\documentclass[twocolumn,showpacs,preprintnumbers,amsmath,amssymb]{revtex4}
\usepackage{graphicx}% Include figure files
\usepackage{dcolumn}% Align table columns on decimal point
\usepackage{bm}% bold math
\usepackage{amssymb}
\usepackage{enumerate}

\newcommand\klinv{K_L \to invisible}

\def\address{\@ifstar{\address@star}%
  {\@ifnextchar[{\address@optarg}{\address@noptarg}}}

\begin{document}

\author{S.N.~Gninenko$^{1}$}
\author{N.V.~Krasnikov$^{1,2}$}
%\author{V.A.~Matveev$^{1,2}$}

\affiliation{$^{1}$ Institute for Nuclear Research of the Russian Academy of Sciences, 117312 Moscow, Russia \\
$^{2}$ Joint Institute for Nuclear Research, 141980 Dubna, Russia}

%\preprint{APS/123-QED}

%\title{Manuscript Title:\\with Forced Linebreak}% Force line breaks with \\

\title{ Invisible  $K_L$ decays as a probe of new  physics }
%and constraints on new physics}
%\title{The  MiniBooNE anomaly and heavy neutrino decay}

\date{\today}% It is always \today, today,
             %  but any date may be explicitly specified
%\date{June 17, 2009}% It is always \today, today,
             %  but any date may be explicitly specified

\begin{abstract}
The  decay $K_L \to invisible$ has never been experimentally tested. In the Standard Model (SM) its branching ratio for the decay into two neutrinos is helicity suppressed and predicted to be $Br(K_L \to \nu \bar{\nu}) \lesssim 10^{-10}$. We consider several natural extensions  of the SM, such as two-Higgs-doublet (2HDM), 2HDM and light scalar, and mirror dark matter models, those main 
feature is that they allow to avoid the helicity suppression factor and lead to an enhanced
$Br(\klinv)$. For the decay $K_L \to \nu \bar{\nu}$  the smallness of the neutrino mass in the considered 2HDM model is explained by  the smallness of the second Higgs doublet vacuum expectation value. The small nonzero  value of the second Higgs isodoublet can arise as a consequence of nonzero quark condensate. 
We show  that taking into account the most stringent constraints from the $K \to \pi  +invisible$ decay, this process could be in the region of $Br(K_L \to invisible) \simeq 10^{-8}-10^{-6}$, which  is  experimentally accessible. In some scenarios the $\klinv$ decay  could still be allowed while the $K \to \pi  +invisible$ decay is forbidden.  The results obtained 
show that the $\klinv$ decay is a clean probe of new physics scales well above 100 TeV, that is complementary to rare $K \to \pi+invisible$ decay,  and provide a strong  motivation for its sensitive search in a near future experiment.
\end{abstract}
% PACS, the Physics and Astronomy
                             % Classification Scheme.
%\keywords{Suggested keywords}%Use showkeys class option if keyword
                              %display desired
\maketitle

\section{Introduction}
 In the Standard Model (SM)  the branching ratios of the   $K^{+} \to \pi^+ + invisible$ and  $K_L \to \pi^0 + invisible$ decays are  predicted  to be 
\cite{buras}
\begin{equation}
 Br(K_L \rightarrow \pi^0 \nu \bar{\nu}) = (2.6 \pm 0.4) \times 10^{-11}  \,,
\label{k+sm}
\end{equation}  
\begin{equation}
 Br(K^+ \rightarrow \pi^+ \nu \bar{\nu}) = (8.5 \pm 0.7) \cdot 10^{-11}\,, 
\label{klsm}
\end{equation}
with the invisible final state represented by neutrino pairs. A strong  comparison between experiment and theory
is possible  due to the accuracy of both the measurements and the SM calculations of these observables. 
A discrepancy would signal the presence of physics beyond the Standard Model (BSM) making  
the precision measurements of these decays an effective probe to  search for it, see e.g. 
\cite{buras,ciri,bryman,kom,Monika,lgl}.

The branching ratio of the $\klinv$ decay  in the SM is predicted to be very small compared to those of  
Eq.(\ref{k+sm}) and Eq.(\ref{klsm}) for $\nu$ masses laying in the  sub-eV region favored by observations of
$\nu$ oscillations \cite{pdg}.   
Indeed, the $K_L$ has zero spin, and it cannot decay into two massless neutrinos, as it contradicts to momentum and angular momentum conservation simultaneously.
%in the $K_L$ rest frame the $\nu$s produced in the decay $K_L \to \nu \overline{\nu} $ fly away in  opposite directions along the same decay axis. 
%If the $\nu$ and $\overline{\nu}$ are massless, the projection of the sum of their spins on this axis equals $\pm$1. The projections of the orbital angular momentum of the $\nu$s on this axis are equal to
%zero. Since in the initial state we have a scalar particle, the process is forbidden. 
For the case of massive $\nu$s
 their spins in the $K_L$  rest frame  must be opposite and, therefore, one of  them is forced to  have the  "wrong" helicity. This results in  the $K_L\to \nu \overline{\nu}$ decay rate being proportional to the $\nu$ mass squared
$\Gamma (K_L\to \nu \overline{\nu}) \propto \Bigl(\frac{m_\nu}{m_{K_L}}\Bigr)^2 \lesssim 10^{-17}$ 
assuming $m_\nu \lesssim 1$ eV. However, if one take the direct experimental upper limit on the $\nu_\tau$ mass $m_{\nu_\tau}< 18.2$ MeV \cite{pdg}, 
the predicted branching ratio, calculated  at the quantum loop level is \cite{marci}
\begin{equation}
Br(K_L\to \nu \overline{\nu}) \simeq 10^{-10} 
\label{klbr}
\end{equation} 
Therefore, an observed $Br(\klinv) \gg 10^{-10}$ would unambiguously signal the presence of BSM physics. 

  The decay $\klinv$ has never been experimentally tested. Since long ago it was recognized that this 
decay	"would be interesting to explore, but its detection looks essentially impossible. New ingenious experimental ideas are required"  \cite{marci}.   
Recently, an approach for performing such kind of experiments by using  the $K^+ n \to  K^0 p$ (or  $K^-  p \to  \overline{K}^0 n$ ) charge-exchange reaction as a source of well tagged $K^0$'s has been reported \cite{Gninenko}. At the same time the first experimental bound $ Br(K_L \to invisible ) \lesssim 6.3 \cdot 10^{-4}$ 
has been set 
from existing experimental data. It has been shown, that 
compared to this limit,  the expected sensitivity of the proposed search is at least two orders of magnitude higher - $Br(K_L \to invisible) \lesssim 10^{-6}$  per $\simeq 10^{12}$ incident kaons. It could be further  improved by utilizing a  more careful design of the experiment, thus making the region  
$Br(K_L \to invisible) \simeq 10^{-8} - 10^{-6}$, or even below,  experimentally accessible \cite{Gninenko}. 

 Being motivated by these considerations,  we discuss in this work several natural extensions of the SM  and show that taking into account the most stringent constraints from the measured $K^+ \rightarrow \pi^+  +invisible$ decay rate, the decay $\klinv$ could occur  at the level $Br(K_L \to invisible)\simeq 10^{-8} - 10^{-6}$. The main feature of the considered models,  that leads to an enhanced branching ratio for $\klinv$, compared to 
 $K^+ \to \pi^+ +invisible$, is that they allow to avoid the helicity suppression factor $\Bigl(\frac{m_\nu}{m_{K_L}}\Bigr)^2$ in the SM, while profiting from its larger phase-space due to the decay into two  light weakly interacting particles. In addition, there might be the case when $\klinv$ could still be kinematically allowed,  while  $K^+ \to \pi^+ + invisible$ is forbidden.
  Additional motivation to search for the $K_L$ (and $K_S)$ invisible decay is related  to precision tests of the $K^0-\overline{K}^0$ system by using the Bell-Steinberger unitarity relation \cite{Gninenko}. This relation connects $CP$ and $CPT$ violation in the  mass matrix to $CP$ and $CPT$ violation in all decay channels of neutral kaons and  is a powerful tool for testing $CPT$ invariance with neutral kaons \cite{adpdg}. The  question  of how much the invisible decays of $K_S$ or $K_L$ can influence the precision of the Bell-Steinberger analysis  still remains open \cite{shub}. All this makes the future searches for this decay mode  very  interesting and complementary to the study of  the $K \to  \pi + invisible$ decays.

\section{ $K_L \rightarrow \nu\bar{\nu}$ decay in model with additional scalar doublet}
Consider now the $K_L \rightarrow \nu\bar{\nu}$ decay in {\it the two-Higgs-doublet model} (2HDM)  with an additional heavy Higgs doublet $H_2$. This type of 2HDM models can introduce flavor-changing neutral currents, provide explanations of the origin of Dark Matter and $CP$ violation , see e.g. Ref.\cite{branco}. 
The  interaction of the heavy isodoublet field $H_2$ with quarks, leptons and the standard Higgs isodoublet $H$ 
leading to the $K_L \rightarrow \nu\bar{\nu}$ decay 
has 
the form
\begin{eqnarray}
L_{int} = h_{2\tau}\bar{L}_{\tau}\tilde{H}_{2}\nu_{\tau_{R}} + h_{2d_Ls_R} \bar{Q}_{1L}H_2 s_R \\ \nonumber 
 + \delta m^2_{HH_2}H^+H_2  + h.c. -  M^2_{H_2} H^+_2H_2   \,, 
\end{eqnarray}
where $L_{\tau} = (\nu_{\tau_{L}}, \tau_{L})$, $Q_{1L} = (u_L, d_L)$,   
$H_2 = (H_2^+, H_2^0)$, $\tilde{H}_2 = ((H_2^0)^*, -(H_2^+)^*)$ and    $h_{2\tau},  h_{2d_Ls_R}$ are Yukawa coupling constants. 
Note that in general the second Higgs isodoublet $H_2$ will have nonzero Yukawa interactions with other quark and lepton fields
%\footnote{
%The general form of the $H_2$ Yukawa interaction is 
%$L_{H_2Yuk} = \sum_{i,j}h_{ijQ_Ld_R}\bar{Q}_{Li}H_2 d_{Rj} +   \sum_{i,j}h_{ijQ_Lu_R}\bar{Q}_{Li}\tilde{H}_2 u_{Rj}  + 
%\sum_{i,j}h_{ijL_L\nu_R}L_{Li}\tilde{H}_2\nu_{Rj} +  \sum_{i,j}h_{ijL_L e_R}L_{Li}H_2e_{Rj} $  , 
%where 
%$Q_{L1} = (u_L, d_L)$,  $Q_{L2} = (c_L, s_L)$, $Q_{L3} = (t_L, b_L)$, $d_{R1} = d_R$, $d_{R2} = s_R$, $d_{R3} = b_R$, 
%$e_{R1} = e_R$, $e_{R2} = \mu_R$, $e_{R3} = \tau_R$,  $\nu_{R1} = \nu_{e_R}$, $\nu_{R2} = \nu_{\mu_R}$, 
%$\nu_{R3} = \nu_{\tau_R}$.}
but 
since we are interested mainly in the $K_L \rightarrow \nu\bar{\nu}$ decay we have written explicitly 
only the Yukawa interactions important 
for us.  
In considered model the neutrinos acquire  nonzero Dirac masses 
$m_{\nu_{\tau}} = h_{2\tau}<H_2>$
due to nonzero vacuum expectation value of the second Higgs isodoublet $H_2$ 
$<H_2> =  \frac{\delta m^2_{HH_2}}{M^2_{H_2}} <H> $  ($<H> = 174~GeV$) 
and  the smallnes of the 
Dirac neutrino masses is a consequence of  the $<H_2>$  smallnes. 
%In our model we can have both the case with Dirac and Majorana masses. 
%For the   neutrino Dirac mass   
The smallnes of $<H_2>$ is due to the assumed large value of  $M_{H_2}$ or(and) small value of 
$\delta m^2_{HH_2}$\footnote{In ref.\cite{Ma} a model with additional Higgs isodoublet interacting 
only with lepton fields was proposed.In this model the neutrino acquire nonzero Dirac masses due to nonzero vacuum 
expectation value of the second Higgs isodoublet that allows to decrease the see saw  scale from $O(10^{15}~GeV$ to 
$O(10^3)~GeV$.}. For instance, for $m_{\nu_{\tau}} = 0.1~eV$, 
$h_{2\tau} = 0.1$ 
and $M_{H_2} = 10^{5}~GeV$
we find that $  \frac{\delta m^2_{HH_2}}{M^2_{H_2}} = 0.6 \cdot 10^{-11}$ 
and  $\delta m^2_{HH_2} = 0.06~GeV^2 $. 
It is interesting to note that  for $\delta m^2_{HH_2} = 0$ 
the $<H_2> = 0$ at classical level but the spontaneous symmetry breaking of $SU_L(3)\otimes SU_R(3)$ 
chiral symmetry in QCD leads to nonzero vacuum expectation values for the Higgs fields\cite{Tokarev}. 
Really, for monzero  Yukawa interaction $L_{H_2Q_1d} = h_{2d_{L}d_{R}} \bar{Q}_{1L} H_2 d_R + h.c~.$    
due to nonzero vacuum expectation value of quark condensate $<\bar{d}d> = - \frac{f^2_{\pi}m^2_{\pi}}{(m_u + m_d)}$ 
($f_{\pi} = 93~MeV$)  
the field $<H_2>$ acquires monzero vacuum expectation value $<H_2 > = \frac{< \bar{d}d >}{ 2 h_{2d_{L}d_{R}}M^2_{H_2}}$.
Numerically for $h_{2\tau} = h_{2d_{L}d_{R}} =1$ and $m_{\nu_{\tau}} = 0.1~eV$ we find that 
$M_{H_2} \sim O(10^4)~GeV$. So in this model with $\delta m^2_{HH_2} = 0$ the vacuum expectation value $<H_2> = 0$ at 
tree level but the nonzero quark condensate leads to the appearance of small vacuum expectation value 
$<H_2> \neq 0$ for the second Higgs isodoublet that explains the smallnes of the neutrino masses. 
 
For the case of nonzero neutrino Majorana mass  $m_{\nu_{\tau_{R}}}$
we assume that the mass $m_{\nu_{\tau_{R}}}$  is 
 small so the decay $K_L \rightarrow \nu_{\tau} \bar{\nu_{\tau}}$ 
is kinematically allowed. 
Again, as in the previous case we assume that the Dirac neutrino mass arises due to nonzero $<H_2>$  
vacuum expectation value and the smallness of the see saw $m_{\nu_{\tau_{R}}} = \frac{m^2_{D\nu_{\tau}}}{m_{\nu_{\tau_{R}}}}$ 
neutrino mass is again explained due to the smallness of $<H_2>$.
The Lagrangian (4) contains $\Delta S = 1$ neutral flavour changing terms but 
for heavy doublet $H_2$ it is not dangerous.
The effective four fermion Lagrangian describing the decay 
 $K_L \rightarrow \nu_{\tau} \bar{\nu_{\tau}}$  has the form 
\begin{equation}
L_{eff} = \frac{1}{M_X^2} \bar{d}_Ls_R\bar{\nu}_{\tau_{L}}\nu_{\tau_{R}} + h.c. \,, 
\end{equation}
where 
\begin{equation}
\frac{1}{M_X^2} = \frac{h_{2d_Ls_R}h_{2\tau}}{M^2_{H_2}} \,.
\end{equation}

As it has been mentioned before we assume the existence of 
small Dirac or Majorana neutrino mass $\nu_{\tau}$. 
The decay rate of the invisible decay $K_L \rightarrow \nu_{\tau}\bar{\nu}_{\tau}$ is 
determined by formula
\begin{eqnarray}
\Gamma(K_L \rightarrow \nu_{L\tau}\bar{\nu}_{R\tau}, \nu_{R\tau}\bar{\nu}_{L\tau}) =
\frac{M^5_{K_L}}{16\pi M^4_X} \nonumber \\
(\frac{F_K}{2(m_d + m_s)})^2K(m^2_{\nu}/M^2_{K_L}) \,,
\end{eqnarray}
where $K(x) = (1-4x)^{1/2}$ for Dirac neutrino with a mass $m_{\nu_{\tau}}$ and  $K(x) = (1 - x)^2$  
for Majorana neutrino $\nu_{\tau_R}$ with a mass $m_{\nu_{\tau_{R}}}$. 
Here  $F_K \approx 160~MeV$ is kaon decay constant 
and $m_s, m_d$ are the masses of $s$- and $d$-quarks\footnote{The quark masses $m_d,m_s$ and the effective 
mass $M_X$ implicitly depend on the renormalization point $\mu$ but  their combination $M^2_X(m_d + m_s)$ 
and hence the decay width (7)
is renormalization 
groop invariant and does not depend on the renormalization point $\mu$.}.   For 
$Br(K_L \rightarrow \nu_{\tau}\bar{\nu_{\tau}}) = 10^{-6}$ we can test the value of $M_{X}$ up to
\footnote{In our estimate (8) we used 
the values $\tau(K_L) = 5.17\cdot 10^{-8}~sec$, $m_{\nu_{\tau}} = 0 $ and $(m_d + m_s)(\mu = 1~GeV) = 160~MeV$}
\begin{equation}
M_{X} \lesssim  0.6 \cdot 10^5~GeV
\end{equation}
for small Dirac or Majorana neutrino mass $m_{\nu_{\tau}} \ll M_{K_L}$. 

It should be noted that the existence of $\Delta S  = 1$ neutral flavour changing interaction (5) leads to 
additional contribution to rare decays $K_L \rightarrow \pi^0 \nu \bar{\nu}$ and 
$K^+ \rightarrow \pi^+ \nu \bar \nu$. The current experimental values are  \cite{PartData}, \cite{PartData1}
\begin{equation}
 Br(K_L \rightarrow \pi^0 \nu \bar{\nu}) < 2.6  \times 10^{-8} \,,
\end{equation}  
\begin{equation}
 Br(K^+ \rightarrow \pi^+ \nu \bar{\nu}) = (17.3^{+11.5}_{-10.5}) \cdot 10^{-11}\,, 
\end{equation}  
with  the SM predictions of (\ref{k+sm}) and (\ref{klsm}), respectively.
%\begin{equation}
% Br(K_L \rightarrow \pi^0 \nu \bar{\nu}) = (2.6 \pm 0.4) \times 10^{-11}  \,,
%\end{equation}  
%\begin{equation}
% Br(K^+ \rightarrow \pi^+ \nu \bar{\nu}) = (8.5 \pm 0.7) \cdot 10^{-11}\,, 
%\end{equation}  
 The measured value (10)  for the  $Br(K^+ \rightarrow \pi^+ \nu \bar{\nu})$   allows to set more  stringent constraints. Therefore, we restrict ourselves to the calculation of the BSM  contribution only to this  decay channel by using
the effective Lagrangian (5). This leads to the following formula 
for the differential  $K^+ \rightarrow \pi^+ \nu \bar{\nu}$ decay width: 
\begin{eqnarray}
\frac{d\Gamma^{BSM}(K^+ \rightarrow \pi^+\nu\bar{\nu})}{dq^2} =\frac{1}{(2\pi)^3}\cdot \frac{1}{32 M^3_{K^+}}\cdot\frac{(q^2 - m^2_{\nu_{\tau,R}})^2}{q^2 M^4_X} & \nonumber \\
 \cdot \sqrt{(M^2_{K^+} + M^2_{\pi^+} - q^2)^2 - 4 M^2_{K^+}M^2_{\pi^+}}  \nonumber \\
\cdot \Bigl[\frac{f_0(q^2)(M^2_{K^+} - m^2_{\pi^+})}{2(-m_d + m_s)}\Bigr]^2~~~~ 
\end{eqnarray}

The form factor $f_{0}(q^2)$ is determined in standard way as
\begin{equation}
<\pi|\bar{d}\gamma_{\mu}s|K> = f_{+}(q^2)(P_K +P_{\pi})^{\mu} + f_{-}(q^2)(P_{K} - P_{\pi})^{\mu} = 
\end{equation}
$$
f_{+}(q^2)[(P_K + P_{\pi})^{\mu} - \frac{M^2_K - M^2_{\pi}}{q^2}q^{\mu}] + 
f_{0}(q^2)\frac{M^2_K - M^2_{\pi}}{q^2}q^{\mu}\,,
$$ 
where $q^{\mu} = (P_K - P_{\pi})^{\mu}$ 
and $ m^2_{\nu_R} \leq (M_{K^+} -M_{\pi^+})^2$. The form factors $f_{+}$ and $f_{0}$ are related to the exchange of $1^{-}$ and $0^{+}$, 
respectively. The following relation holds:
\begin{equation}
f_{+}(0) = f_0(0)~, ~f_0(q^2) = f_+(q^2) + \frac{q^2}{M^2_K - M^2_{\pi}}f_{-}(q^2) \,.
\end{equation} 
In our calculations we use standard  linear parametrization 
for the form factor $f_0(q^2)$, namely 
\begin{equation}
f_{0}(q^2) = f_0(0)(1  +\lambda_0\frac{q^2}{M^2_{\pi^+}})\,. 
\end{equation} 
Numerically  we take  $f_0(0) = 0.96$\cite{Form} and $\lambda_0 = -0.06$ \cite{Form1}.
 
It is convenient to represent the result in terms of the ratio 
$\beta^{-1} \equiv \frac{Br(K_L  \rightarrow  \nu \bar{\nu})}{Br(K^+  \rightarrow  \pi^+ \nu \bar{\nu})}$
because the ratio $\beta$ 
does not depend on unknown value of $M_X$. Also $\beta$ does not depend on the values of quark masses $m_d, ~m_s$. 
For the case of massless neutrino we find that 
\begin{equation}
\beta  \approx 2 \cdot 10^{-3} \,.
\end{equation}
Note that the smallness of the $\beta$ is mainly due to the 3-body phase space smallnes  in comparison 
with 2-body phase space. 
From the difference between the theoretical and experimental values (2) and (10), respectively,  
by summing up errors of (10) in quadrature we find  
that the BSM contribution to the $ Br(K^+ \rightarrow \pi^+ \nu \bar{\nu})$
is less than 
\begin{equation}
Br^{BSM}(K^+ \rightarrow \pi^+ \nu \bar{\nu}) \lesssim  2.1 \cdot 10^{-10} \,. 
\end{equation}
From the limit (16) and the estimate (15) we find that  for massless neutrinos 
\begin{equation}
Br(K_L \rightarrow \nu \bar{\nu}) \lesssim  10^{-7}
\end{equation}
The estimates (15, 17)  are valid for small $m_{\nu_R} \ll M_{\pi^+}$ Majorana msss of 
righthanded neutrino. For higher  $m_{\nu_R}$ values  the limit (17) is more weak and 
for the case  $M_{K_L} \geq      m_{\nu_R} \geq M_{K^+} -M_{\pi^+}$  when the decay 
$K^+ \rightarrow \pi^+ \nu_{\tau_L} \bar{\nu}_{\tau_R}$ is kinematically prohibited, but the decay 
$K_L \rightarrow \nu \bar{\nu}$ is still allowed, the restriction from $K^+ \rightarrow \pi^+ \nu \bar{\nu}$ decay  does not work. 

The measured $(K_L - K_S)$ mass difference strongly restricts \cite{Batler} the effective  $\Delta S = 2$ interaction 
\begin{equation}
L_{\bar{s} d\bar{s}d} = \frac{1}{\Lambda^2_{\bar{s}d\bar{s}d}} \bar{s_R}d_L\bar{s_R}d_L ~+~h.c.~ \,.
\end{equation}
Namely \cite{Batler}  
\begin{equation}
\Lambda_{{\bar{s}d\bar{s}d}} \geq 1.8 \cdot 10^{7}~GeV \,.
\end{equation}
For the model (2) with the the additional Higgs doublet 
$H_2 = (H_{2}^+, H^0_{2,1} + i H^0_{2,2})$ we find that
\begin{equation}
\frac{1}{\Lambda^2_{\bar{s}d\bar{s}d}}  = 
|h_{2d_Ls_R}|^2|\frac{1}{M^2_{H^0_{2,1}}} - \frac{1}{M^2_{H^0_{2,2}}}| \sim
\frac{|h_{2d_Ls_R}|^2}{M^2_{H_2}}\cdot \frac{\delta{m^2_{HH_2}}}{M^2_{H_2}} \,.
\end{equation}
Using the bound (19) we can restrict  the 
parameter $ {\delta{m^2_{HH_2}}}$. For instance, for $M_{H_2} = 10^{5}~GeV $, $h_{2d_Ls_R} = 1 $ 
we find $ {\delta{m^2_{HH_2}}} \leq 0.3 \cdot 10^6~ GeV^2$ that is much more weak than the estimate of 
  $\delta m^2_{HH_2}$ coming from the neutrino mass. 

In general case we can have additional flavour changing Yukawa interaction $h_{2s_Ld_R}\bar{Q}_{2L}H_2d_R + h.c$ ($Q_{2L} = (c_L,s_L)$ in 
the Lagrangian (4) that leads to the tree level flavour changing $\Delta S = 2$ effective interaction $
L_{eff} = \frac{h_{2d_Ls_R}h^*_{2s_Ld_R} }{M^2_{H2}}(\bar{d}_Ls_R\bar{d}_Rs_L + h.c.)$ 
We can simultaneously avoid the $\Delta S = 2$ bound $ \Lambda_{\Delta s = 2} \equiv  (h_{2d_Ls_R}h^*_{2s_Ld_R})^{-1/2} \cdot M_{H_2} 
 > 1.8 \cdot 10^{7}~GeV$  and obtain phenomenologically interesting values for $Br(K_L \rightarrow \nu \bar{\nu})$ for 
small quark Yukawa coupling constants $h_{2d_Ls_R},~h_{2s_Ld_R}$, relatively light  second Higgs doublet  
and not small lepton Yukawa coupling constant $h_{2\tau}$ . For instance, for  $h_{2d_Ls_R} = h_{2s_Ld_R} = (1/300)^2$ , 
$h_{2\tau} = 1$ and $M_{H_2} = 300~GeV$  we find that   $\Lambda_{\Delta s = 2} = 2.7 \cdot 10^{7}~GeV $ and 
$Br(K_L \rightarrow \nu \bar{\nu}) =  0.4 \cdot10^{-6} $. The existence of relatively light 
with a mass $M_{H_2} = 300~GeV$  second Higgs doublet does not contradict the LHC data. The best way to look for 
the second Higgs isodoublet at the LHC is the use of the reaction $pp \rightarrow Z^*/gamma^* \rightarrow H^+_2H^{-}_2 
\rightarrow \tau^+\tau^{-} \nu \bar{\nu}$. So the signature is two  $\tau$ leptons plus nonzero $E^T_{miss}$ in 
final state that coincides with the signature used for the search for direct production of stau leptons at the LHC.

%Note that in this section we considered only the parts of the Yukawa interactions which responsible for 
%nonzero $K_L \rightarrow \nu \bar{\nu}$ decay. The second Higgs isodoublet $H_2$ can have nonzero interactions 
%with other quark and leptons that can lead to other invisible   %$D^0,B^0 \rightarrow \nu\bar{\nu}$. The 
%existing experimental data restrict but not very strongly in comparison with the 
%%$K^+ \rightarrow \pi^+ \nu\bar{\nu}$ the corresponding Yukawa coupling constants. 

%\footnote{The bound on $\delta m^2_{HH_2}$ from the$K_L - K_S$ mass difference is  much more weak
%  than the estimate on  $\delta m^2_{HH_2}$ coming from the neutrino mass. For instance, for 
%$\nu{\tau} = 0.1~eV$ $h_{2\tau} = 0.1$, $M_{H_2} = 10^5~GeV$ the value of $\delta m^2_{HH_2}$ is predicted equal to 
%$\delta m^2_{HH_2} = 0.06~GeV^2$}.
%Note that in the model  (4) the contribution to the $K_L -K_S$ mass difference is expected to be  
%proportional to $\frac{1}{M^2_{H_2}}\cdot \frac{|<H>|^2}{M^2_{H_2}}$ and it is negligible.  

\section{ $K_L \rightarrow \phi \phi$ decay in model with additional scalar doublet and scalar singlet $\phi$}
Consider now the $K_L \rightarrow \phi \phi$ decay in  the extension of the SM {\it with heavy Higgs doublet $H_2$ and light neutral scalar singlet field $\phi $}.
The Yukawa interaction of the heavy isodoublet $H_2$ with quarks and the interaction of the $\phi$ field 
with Higgs isodoublets $H_2$ and $H$(Higgs  isodoublet of the SM) has the form 
\begin{eqnarray}
L_I = h_{2d_Ls_R}\bar{Q}_{1L}s_RH_{2}  + \lambda (H_2^+H)\phi^2 +  \delta m^2_{HH_2}H^+H_2\\ \nonumber
+ h.c. -  M^2_{H_2} H^+_2H_2  \,,
\end{eqnarray}
where $Q_{1L} = (u_L, d_L)$, $H_2 = (H^+_2, H^0_2)$  
and $h_{2d_Ls_R}$, $\lambda$ are Yukawa and Higgs couplings. After electroweak $SU_L(2)\otimes U(1)$ symmetry breaking trilinear term 
describing transition $H_2 \rightarrow \phi\phi$
\begin{equation}
L_{H_2\phi\phi} = \lambda <H>H_2^+\phi^2   + h.c. \,
\end{equation}
arises. The effective Lagrangian
\begin{equation}
L_{eff} = \frac{1}{M_X} \bar{d}_Ls_R \phi^2 + h.c. \,, 
\end{equation}
\begin{equation}
\frac{1}{M_X} = \frac{h_{2d_Ls_R}\lambda <H>}{M^2_{H_2}} \, 
\end{equation}
describes invisible decay $K_L \rightarrow \phi\phi$. Here we assume that the 
mass of $ \phi$ is less than $M_{K_{L}}/2$. 
The decay rate of the invisible decay $K_L \rightarrow \phi \phi$ is 
determined by formula
\begin{equation}
\Gamma(K_L \rightarrow \phi \phi ) =
\frac{M^3_{K_L}}{8\pi M^2_X}(\frac{F_K}{2(m_d + m_s)})^2K(m^2_{\phi}/M^2_{K_L}) \,,
\end{equation}
where $K(x) = (1-4x)^{1/2}$.  
For  $Br(K_L \rightarrow \phi \phi) = 10^{-6}$ and $m_{\phi} \ll M_{K_L}$  we can test the value of $M_{X}$ up to
\begin{equation}
M_{X} \lesssim   10^{10}~GeV \,.
\end{equation}
For $\lambda =1$ and  $h_{2d_Ls_R} = 1$  the mass of the second Higgs isodoublet can be tested up to $M_{H_2} \leq 10^{6}~GeV$.

The bound  (16) allows to 
restrict the $K_L \rightarrow \phi \phi $ decay  in full analogy with previous model. 
Namely, in the model with the effective Lagrangian (22) the $K_L \rightarrow \phi \phi $ 
decay width is determined by the expression
\begin{eqnarray}
\frac{d\Gamma^{BSM}(K^+ \rightarrow \pi^+\phi \phi)}{dq^2} = \frac{1}{(2\pi)^3}\cdot \frac{1}{32 M^3_{K_L}}\cdot\frac{2}{M^2_X} & \nonumber \\
 \cdot \sqrt{[(M^2_{K^+}+M^2_{\pi^+}-q^2)^2-4 M^2_{K^+}M^2_{\pi^+}]}(1-\frac{4m^2_{\phi}}{q^2})\cdot &\nonumber \\ 
\Bigl [\frac{f_0(q^2)(M^2_{K^+} - m^2_{\pi^+})}{2(-m_d + m_s)} \Bigr ]^2.~~~
\end{eqnarray}

It is convenient to use  the ratio 
$\beta^{-1} \equiv \frac{\Gamma(K_L  \rightarrow  \phi \phi)}{\Gamma(K^+  \rightarrow  \pi^+ \nu \bar{\nu})}$
because the ratio $\beta$ 
does not depend on unknown value of $M_X$ and  on the values of quark masses $m_d, ~m_s$. 
For the case  $m_{\phi} \ll M_{\pi^{+}}$ 
we find that 
\begin{equation}
\beta  \approx  10^{-2} \,.
\end{equation}
As in the previous model  the smallness of the $\beta$ is mainly due to the 3-body phase space smallnes  in comparison 
with 2-body phase space. 

From the (16) and (28) we find  
\begin{equation}
Br(K_L \rightarrow \phi \phi) \lesssim 2 \cdot 10^{-8} \,.
\end{equation}
For not very light $\phi$-particle the limit on $Br(K_L \rightarrow \phi \phi)$ will be not so 
stringent as the bound
(29), moreover, for $\phi$  particle mass $ M_{K_L}/2      \geq  m_{\phi} \geq (M_{K^{+}} - M_{\pi^+})/2$
 the decay $K^+ \rightarrow \pi^+ \nu \bar \nu$ is kinematically prohibited while the decay 
$K_L \rightarrow \phi \phi $ is allowed.  Therefore  the  bound (29)
derived from the decay width of the $K^+ \rightarrow \pi^+ \nu \bar \nu$ decay 
does not work for $K_L \rightarrow \phi \phi$ decay mode. Note, that such  sub-Gev  scalar $\phi$ could be a good dark matter candidate \cite{Fayet}. 
As in the previous model the bound from the $K_L - K_S$ mass difference leads to the bound on the unknown parameter $ {\delta{m^2_{HH_2}}}$ 
at the level  $ {\delta{m^2_{HH_2}}} \leq 30~ GeV^2$ for  $\frac{M_{H_2}}{h_{2d_Ls_R}} = 10^{4}~GeV $.

\section{$K_L \rightarrow invisible$ decay in model with mirror world}

Finally, we discuss  the $K_L $ oscillations into a hidden sector, which would manifest themselves through 
the $\klinv$ decay.  As an example of such
hidden sector we consider the one of {\it the mirror matter models}.
The idea that along with the ordinary matter may exist
its exact mirror copy, introduced  for the parity conservation, 
is not new  \cite{mirror}. Accordingly, each ordinary
particle of the SM has a corresponding mirror partner of
exactly the same mass as the ordinary one. The mirror fields  are all
singlets under the SM $SU_c(3)\otimes SU_L(2)\otimes U(1)$ 
gauge group. Mirror matter
is dark  in terms of the SM interactions, and could be a
good candidate for dark matter, see, e.g., Refs.\cite{mirrormatter}, and recent  \cite{footdm}.  
In addition to gravity, the interaction between our
and this type of dark matter could be transmitted by some gauge
singlet particles interacting with both sectors. 
Any neutral, elementary or composite particle, in principle, 
can have mixing with its mirror duplicate. This results in several interesting phenomena,
such, e.g. as Higgs \cite{wil}, positronium \cite{pos}, muonium \cite{muonium}, or neutron \cite{zurab}
 oscillations into their hidden partner,  which have been or planned to be  experimentally tested \cite{bader, psinn, ser, paolo}.
   
In particular, the neutral $K_L$-meson can mix with it 
mirror ($m$)  analog $K_{L,m}$ due to effective 
four-fermion interaction 
\begin{equation}
L_{int} = \frac{1}{M^2_{m}}[\bar{d}\gamma^{\mu}(1-\gamma_5)s \bar{s}_{m}\gamma_{\mu}(1-\gamma_5)d_{m}]
\end{equation}
The
interaction (30) leads to conversion of ordinary $K_L$-meson 
to mirror $K_L$-meson. The decays of mirror  $K_L$-meson are invisible in our world that 
leads to invisible $K_L$ decay with the branching ratio 
\begin{equation}
Br(K_L \rightarrow invisible) = \frac{\delta^2}{2(\delta^2 +\Gamma_{tot}^2(K_L))}\,,
\end{equation}
where 
\begin{equation}
\delta = \frac{1}{M_{K_L}}<K_{L,m}|L_{int}|K_L>\,.
\end{equation}
For the interaction (27) in the vacuum insertion approximation 
we find that  
\begin{equation}
\delta \approx  \frac{F^2_K M_{K_L}}{M^2_{m}}  \,.
\end{equation}
Numerically, for $Br(K_L \rightarrow invisible) = 10^{-6}$
we can probe the value of $M_{m}$ up to 
\begin{equation}
M_{m} \lesssim  8.4 \cdot10^8~GeV \,.
\end{equation}
In our estimates we used nonrenormalizable effective four-fermion interaction (30).
It is possible to obtain the effective interaction (27) from renormalizable mirror world model 
with the Higgs doublet extension of the SM model (see previous discussions) and with the additional interaction term between our and mirror world 
\begin{equation}
L_{m} = \lambda_{m}(H^+H_2)(H_{m}H^{+}_{m,2}) + h.c. \,.
\end{equation}
After electroweak symmetry breaking in our and mirror worlds $(<H> = <H_{m}> \approx 174~GeV)$    
we find an effective four-fermion interaction
\begin{equation}
L_{eff} = \frac{1}{M^2_{m}}\bar{d}_Ls_R\bar{s}_{R,m}d_{L,m} +h.c. \,,
\end{equation}
where 
\begin{equation}
\frac{1}{M_{m}^2} = \frac{h_{2d_Ls_R}^2}{M^2_{H_2}}\cdot \frac{\lambda_{m}|<H>|^2}{M^2_{H_2}} \,. 
\end{equation}

\section{Conclusion}
 In conclusion, the observation of the  $K_L \to invisible$ decay with the branching ratio 
$Br(\klinv) \gg 10^{-10}$ would unambiguously signal the presence of BSM physics. We consider   the $K_L \to invisible$ decay in several natural extensions of the SM, such as the 2HDM, 2HDM and light neutral scalar field $\phi $, and mirror dark matter model.  Using constraints from the experimental value for  the  $Br(K^+ \rightarrow \pi^+ \nu \bar{\nu})$ we find that the  $K_L \to invisible$ decay  branching ratio could be in the   region $Br(K_L \to invisible)\simeq 10^{-8}- 10^{-6}$, 
which is  experimentally accessible allowing to test new-physics scales well above 100 TeV. In some scenarios
 these bounds can be avoided, as  in the model with the massive  right-handed neutrino and scalar 
$\phi$-particle. This makes  the  $K_L \to invisible$ decay a  powerful clean  probe of  new physics, that is complementary to  other  rare $K$ decay channels. Additionally, in the case of observation  the  $K_L \to invisible$ decay could  influence the precision of the Bell-Steinberger analysis of the $K^0-\overline{K}^0$ system. The results obtained provide a strong motivation for a sensitive search for this process in a near future $K$ decay experiment proposed in \cite{Gninenko}. It should be noted that in 
full analogy with the case of $K_L$ invisible decay we can  expect the existence of 
invisible decays of $B_d$ and $B_s$ mesons, see e.g. \cite{alex,babar}, 
 with the branchings similar to those discussed above.

This work was supported by the Grant RFBR N 13-02-00363.

\end{document}